\documentclass[data,datadescriptor,accept,moreauthors,pdftex,10pt,a4paper]{mdpi} 
\firstpage{1} 
\articlenumber{8}
\doinum{10.3390/data3010008}
\pubvolume{3}
\pubyear{2018}
\copyrightyear{2018}
\history{Received: date; Accepted: date; Published: date}

\pdfoutput=1

\preto{\abstractkeywords}{\nolinenumbers}

\usepackage{booktabs} 
\usepackage{multirow}
\usepackage{soul} 
\usepackage{microtype}
\usepackage{array}
\usepackage{upgreek}
\Title{RAE: The Rainforest Automation Energy Dataset for Smart Grid Meter Data Analysis}

\Author{Stephen Makonin $^{1,}$*\orcidA{}, Z. Jane Wang $^{1}$ and Chris Tumpach $^{2}$}

\AuthorNames{Stephen Makonin, Z. Jane Wang and Chris Tumpach}

\address{%
$^{1}$ \quad Electrical and Computer Engineering, University of British Columbia, Vancouver, BC, Canada, V6T 1Z4; smakonin@alumni.ubc.ca, zjanew@ece.ubc.ca\\
$^{2}$ \quad Rainforest Automation, Inc., Burnaby, BC, Canada, V5G 4P5; chris.tumpach@rainforestautomation.com}

\corres{Correspondence: smakonin@alumni.ubc.ca}%; Tel.: +x-xxx-xxx-xxxx}

\abstract{Datasets are important for researchers to build models and test how well their machine learning algorithms perform. This paper presents the Rainforest Automation Energy (RAE) dataset to help smart grid researchers test their algorithms that make use of smart meter data. This initial release of RAE contains 1 Hz data (mains and sub-meters) from two residential houses. In addition to power data, environmental and sensor data from the house\rq{}s thermostat is included. Sub-meter data from one of the houses includes heat pump and rental suite captures, which is of interest to power utilities. We also show an energy breakdown of each house and show (by example) how RAE can be used to test non-intrusive load monitoring (NILM) algorithms.}

\keyword{dataset; residential; smart grid; smart meter; in-home display; IHD; HEMS; non-intrusive load monitoring; disaggregation; NILM}

\dataset{doi:\href{http://dx.doi.org/10.7910/DVN/ZJW4LC}{10.7910/DVN/ZJW4LC}}
\datasetlicense{CC-BY}%license under which the data set is made available (CC0, CC-BY, CC-BY-SA, CC-BY-NC, etc.)}

\begin{document}
%%%%%%%%%%%%%%%%%%%%%%%%%%%%%%%%%%%%%%%%%%
%% Only for the journal Gels: Please place the Experimental Section after the Conclusions

%%%%%%%%%%%%%%%%%%%%%%%%%%%%%%%%%%%%%%%%%%
% A short summary of the data set, methods, background information on why and how the data set was collected, short description of funded or unfunded research projects that are or will eventually be based on the data set, and list of publications based on the data set that were possibly already published. Optionally, authors may wish to describe potential benefits of publicly releasing and describing the dataset. In general, the Summary section is similar to an introduction section in a research article.
\section{Summary}

Datasets are becoming increasingly more relevant when measuring the accuracy of smart grid algorithms and seeing how well they might perform in a real-world situation. Testing the accuracy performance with real-world datasets is crucial in this field of research. Synthesized data does not realistically represent an actual dataset as ``a real-world dataset would normally have certain complexity that is harder to predict and in many cases can be very difficult to deal with'' \mbox{\cite{hadzic2011mining} (p. 114)}. For~smart grid research, it is valuable to have public datasets that show how smart meters report aggregate power readings with the accompanying sub-meter data for the different loads that comprise that aggregate reading. This is very true when testing non-intrusive load monitoring (NILM) algorithms~\cite{hart1992nilm,makonin2016tsg}. NILM (sometimes referred to as load disaggregation) is a computational approach to determining what appliances are running in a given house (or building) and only involves examining the aggregate power signal from a smart meter.

For the initial release of the RAE dataset, we consider two houses: House 1 and House 2. We are actively assessing other houses that can be monitored and added to this dataset. The~monitoring system that we present here is an accurate and reliable data capture system that can be easily installed in a house to collect data in the same format and frequency. Researchers interested in installing this system and adding data to RAE can contact the lead author.

In addition to smart grid and NILM, this dataset can be used in research that looks at statistical signal processing and blind source separation, energy use behaviour, eco-feedback and eco-visualizations, application and verification of theoretical algorithms/models, appliance studies, demand forecasting, smart home frameworks, grid distribution analysis, time-series data analysis, energy-efficiency studies, occupancy detection, energy policy and socio-economic frameworks, and~advanced metering infrastructure (AMI) analytics. 

\subsection{Relation to Prior Datasets}

Previously, we created a widely used dataset, named the \textit{Almanac of Minutely Power dataset} (AMPds1~\cite{makonin2013ampds} and AMPds2~\cite{makonin2016ampds2}), which contained data sampled at 1 min intervals. This new dataset has all power panel circuits sampled at 1 Hz. Besides AMPds and this dataset, and at the time of writing this, there are no other Canadian open public datasets.

One of the first and well-known datasets, the \textit{Reference Energy Disaggregation Data Set} (REDD)~\cite{kolter2011redd}, which was released in 2011 (USA homes), has a low-frequency sampling version where the mains are sampled at a frequency (1 Hz, or per second) that is higher than the sub-metered loads (per 3 s). It is worth noting that a more recent dataset, called the \textit{UK Domestic Appliance-Level Electricity} (UK-DALE) dataset~\cite{kelly2015ukdale}, employs this methodology as well. The RAE dataset has a different approach. 
The lower the sampling frequency, the more signal features missed at capture. Therefore, it is best to sample the sub-metered loads at a higher sampling frequency so that interesting features from the appliance\rq{}s power signature can be captured. Further, we wanted the mains data to be sampled at a sampling frequency that is common to most smart meter in-home displays (e.g., Rainforest Automation\rq{}s EMU2).

The aforementioned datasets (in the area of NILM) are considered low-frequency sampling ($\leq$1~Hz) datasets. There are indeed high-frequency sampling datasets. REDD does have a high-frequency version of its data. Two such examples are the \textit{Building-Level fUlly labeled Electricity Disaggregation} dataset (BLUED)~\cite{filip2011blued}, sampled at 12 kHz (USA data), and the \textit{Controlled On/Off Loads Library} dataset (COOLL)~\cite{picon2016cooll}, sampled at 100 kHz (France data). While these datasets provide valuable data for high-resolution applications, we feel that it is a more realistic scenario to use low-frequency sampled data for most smart grid and NILM systems, especially where there is a processor constraint on storage and speed.

%%%%%%%%%%%%%%%%%%%%%%%%%%%%%%%%%%%%%%%%%%
% What data is contained? Which format? How can it be read and interpreted? For example, in tabular data give a full description of each column heading.
\section{Data Description}

\begin{figure}[H]
\centering
\includegraphics[width=15.3cm]{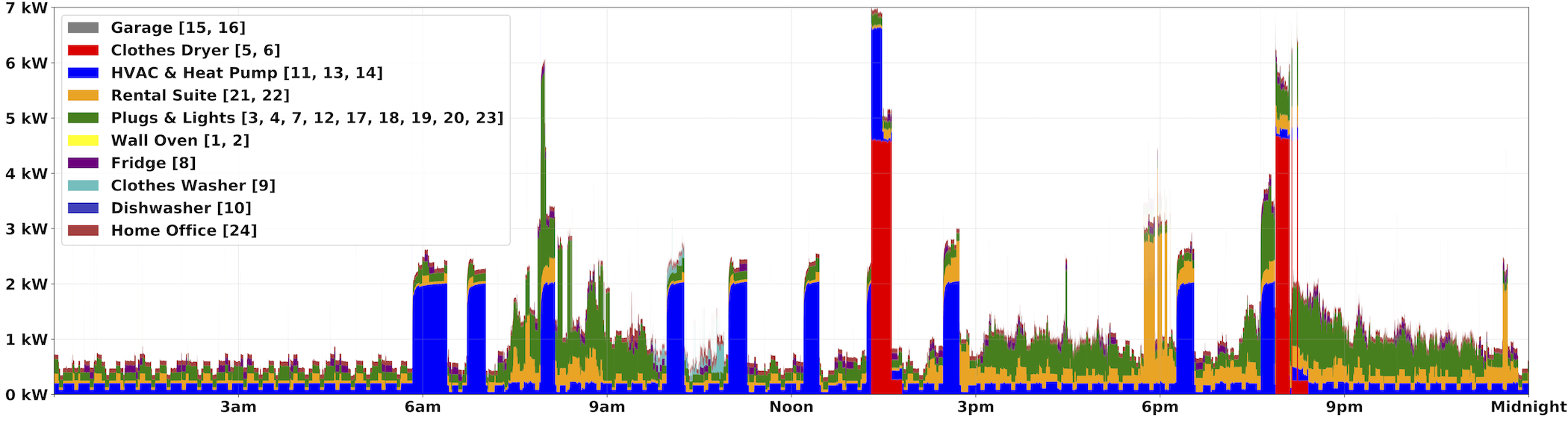}
\caption{Plot of all loads over 24 h on Sunday, March 20, 2016 for House 1.}
\label{fig:day}
\end{figure}

 This dataset contains over 11.3 million power readings. There are up to 24 sub-meters (one for each breaker on the house\rq{}s main power panel) sampled at 1 Hz, which capture 11 electrical data-points (voltage, current, frequency, power factor, real power/energy, reactive power/energy, and apparent power/energy). There are 72 days of capture for House 1 and 59 days for House 2. We also included readings for an \textit{in-home display} (IHD), which samples as a typical ``smart meter communication to in-home display''-rate (per 8--15 s). For House 1, this results in roughly 414,000 samples over the 72~days of capture. 
 By providing IHD data, researchers can gain valuable insight as to how data is given to occupants compared to a constant 1 Hz data stream. We also include environmental and sensor data from the house\rq{}s thermostat, which further augments the understanding of HVAC consumption. Figure~\ref{fig:day} depicts an arbitrary Sunday (a 24 h period) to give the reader a visual idea of what the load consumption pattern can look like.

This dataset has two overall files, \texttt{all\_sites.txt} and \texttt{all\_types.txt}, and a number of site-specific data files which are described in Table~\ref{tbl:files}. 
The file \texttt{all\_sites.txt} contains summary information on all the monitored sites in the dataset. A house would be considered a monitoring site. As different monitoring sites are added, the type of sites will be defined in the \texttt{all\_types.txt} file.

\begin{table}[H]
  \caption{Dataset file descriptions.}
  \centering
\begin{tabular}{m{4cm}<{\centering}m{10cm}}
  \toprule
	\textbf{File Name} & \multicolumn{1}{c}{\textbf{Description}} \\
  \midrule
	 \texttt{all\_sites.txt}          &  Summary data for all monitored sites (e.g., houses). See Table~\ref{tbl:allh} for a~description of the columns in this file.\\\midrule
	 \texttt{all\_types.txt}          &  A dictionary that describes the type of sites that were monitored.\\\midrule
	 \texttt{<type>?.txt}               &  Metadata for the given \textit{<type>} of site monitored followed by its ID. For example, a house of ID 1 would have the filename \texttt{house1.txt}. See Table~\ref{tbl:hmeta} for more details. There is one file for each site.\\\midrule
	 \texttt{<type>?\_energy\_blk?.csv} &  Energy data recorded at hourly intervals for all sub-meters. See~Table~\ref{tbl:power} for more details. There is one file for each reading block of each site.\\\midrule
	 \texttt{<type>?\_labels.txt}       &  Descriptions of each sub-meter monitored (the sub-meter number followed by description), one per line. The number corresponds to the \textit{sub} column in the power and energy data files. (e.g., 1 would be \texttt{sub1}). There is one file for each site.\\\midrule
	 \texttt{<type>?\_panel.pdf}        &  A diagram of the power panel of each house showing the layout fo the breakers and what breakers where monitored. There is one file for each~site.\\\midrule
	 \texttt{<type>?\_power\_blk?.csv}  &  Power data recorded at 1 Hz for all sub-meters. IHD data is also recorded but appears at a lower frequency. See Table~\ref{tbl:power} for more details. There is one file for each reading block of each site.\\\midrule
	 \texttt{<type>?\_subs\_blk?.csv}  &  Extensive electrical measurements for all sub-meters. See Table~\ref{tbl:subs} for a list of these measurements. There is one file for each reading block of each site.\\\midrule
	 \texttt{<type>?\_tstat\_blk?.csv}  &  HVAC thermostat data recorded at approximately 5 min intervals for each thermostat in a house. Data in these files are highly diverse and depend on the thermostat make/model. To compensate for this, columns in these files are verbosely named. There is one file for each reading block of each site.\\
  \bottomrule
  \end{tabular}
  \label{tbl:files}
\end{table}
\vspace{-12pt}

\begin{table}[H]
  \caption{Column descriptions for the \texttt{all\_sites.txt} file.}
  \centering
\begin{tabular}{m{3cm}<{\centering}m{11cm}}
  \toprule
	\textbf{Column Name} & \multicolumn{1}{c}{\textbf{Description}} \\
  \midrule
	 \texttt{type}         &  The type of site monitored. For example, \textit{house} would mean residential and could be detached, row, or apartment. Future values could include \textit{store}, for a store front, \textit{industry}, for an industrial complex, \textit{office}, etc. See the \texttt{all\_types.txt} file.\\\midrule
	 \texttt{id}           &  The house/store/etc. ID number, starting at 1.\\\midrule
	 \texttt{power\_data}  &  An indicator of whether power and energy data is available (Yes/No). Power data is usually recorded at 1 Hz, whereas energy data is recorded in hourly intervals.\\\midrule
	 \texttt{submeters}    &  The number of power sub-meters monitored.\\\midrule
	 \texttt{tstat\_data}  &  An indicator of whether HVAC thermostat data is available (Yes/No).\\\midrule
	 \texttt{block\_count} &  The number of contiguous recorded data blocks for the given house. \\\midrule
	 \texttt{timezone}     &  The timezone in which the given house is located. \\\midrule
	 \texttt{active}       &  An indicator of whether this house is still under active monitoring (Yes/No). If so, more house data will be added in the future.\\
  \bottomrule
  \end{tabular}
  \label{tbl:allh}
\end{table}

\begin{table}[H]
  \caption{Metadata description files for each house.}
  \centering
\begin{tabular}{m{2.5cm}<{\centering}m{11cm}}
  \toprule
	\textbf{Column Name} & \multicolumn{1}{c}{\textbf{Description}} \\
  \midrule
	 <Site Type> ID  &  The ID number of the monitored site. If the site is a house then the row heading will read \textit{House ID}.\\\midrule
	 Type Details    &  A description of the monitoring site. \\\midrule
	 Location        &  The city, province, and country in which the monitored site is located.\\\midrule
	 Local Timezone  &  The local timezone of the monitored site.\\\midrule
	 Year Built      &  The year that the building was built.\\\midrule
	 Year Last Reno  &  The last year that any major renovations were made.\\\midrule
	 EnerGuide       &  If the building has an EnerGuide rating, when it was given.\\\midrule
	 HVAC Type       &  A description of the type of HVAC system installed at the monitored site.\\\midrule
	 Lighting        &  A description of the type of lighting used at the monitored site.\\\midrule
	 Thermostat(s)   &  A list of HVAC thermostats on site, including their make and~model.\\\midrule
	 IHD Device      &  The model of Rainforest Automation in-home display used to record smart meter~data.\\\midrule
	 Sub-meter Equip &  The model of equipment used to monitor power panel breakers.\\\midrule
	 Sub-meter Count &  The number of sub-meters/breakers monitored.\\\midrule
	 Sub-meter Mains &  The aggregated total power/energy. If value \texttt{calc} is given, then mains is calculated by a summation of all sub-meters. Else, listed are sub-meters that monitored the mains. 
	 For example, \texttt{sub1, sub2} would mean that sub-meter 1 (on L1) + sub-meter 2 (on L2) monitored the mains.\\\midrule
	 Active Site     &  An indicator of whether the site is still being monitored. If so, more data will be added to the dataset for this site in the future.\\\midrule
	 Other DOI/URL   &  A URL for a website with more information about the site. There may be other publications.\\\midrule
	 Floors          &  The number of floors at the site. This is followed by one line per floor. The name of the floor, the area/size of the floor, and the number of occupants that usually inhabit that floor.\\\midrule
	 Occupant Notes  &  The number of special occupancy notes.\\\midrule
	 Sampling Blocks &  The number of contiguous monitoring blocks.\\\midrule
	 Missing Data    &  The number of places where missing data has occurred.\\
  \bottomrule
  \end{tabular}
  \label{tbl:hmeta}
\end{table}

\begin{table}[H]
  \caption{Column descriptions for power and energy data files.}
  \centering
\begin{tabular}{m{2.5cm}<{\centering}m{11cm}}
  \toprule
	\textbf{Column Name} & \multicolumn{1}{c}{\textbf{Description}} \\
  \midrule
	 \texttt{unix\_ts} &  The Unix timestamp is UTC. Note that the local timezone is noted in the house metadata file and \texttt{all\_houses.txt} file.\\\midrule
	 \texttt{ihd}      &  The value reported by the IHD and the given timestamp. An empty (or null) value would means there was no reading given at that timestamp.\\\midrule
	 \texttt{mains}    &  Values in this column are calculated either by a summation of all the sub-meters or by the summation of one or two specific sub-meters used to monitor the mains. This is described in the metadata file for each house.\\\midrule
	 \texttt{sub?}     &  Each sub-meter will have a column from 1 to the number of sub-meters (e.g.,~\texttt{sub1}, \texttt{sub2}, ..., \texttt{sub24}).\\
  \bottomrule
  \end{tabular}
  \label{tbl:power}
\end{table}
\vspace{-12pt}

\begin{table}[H]
  \caption{Measurements captured by the DENT PowerScout 24.}
  \centering
  \begin{tabular}{ccc}
  \toprule
 	\textbf{Column} & \textbf{Description} & \textbf{Units}\\
  \midrule
	0	& Unix Timestamp (since Epoch)	& s\\
	1	& Sub-meter ID (sub)				& $number$\\
	2	& Voltage (V)				    & V\\
	3	& Frequency ($f$)				& Hz\\
	4	& Current (I)					& A\\
	5	& Displacement Power Factor (dPF)& $ratio$\\
	6	& Apparent Power Factor (aPF)	& $ratio$\\
	7	& Real Power (P)					& W\\
	8	& Reactive Power (Q)				& VAR\\
	9	& Apparent Power (S)				& VA\\
	10	& Real Energy (Pt)				& Wh\\
	11	& Reactive Energy (Qt)			& VARh\\
	12	& Apparent Energy (St)			& VAh\\
  \bottomrule
  \end{tabular}
  \label{tbl:subs}
\end{table}

Each house has a \texttt{labels} file to describe the loads that each sub-meter monitored accompanied by a~\texttt{panel} file to depict the house\rq{}s power breaker panel that was sub-metered. Given that these houses are located in Canada, there are larger appliances (e.g., clothes dryers) that have two lines (or~sub-meters) for monitoring (L1 and L2) a single appliance. To combine these two lines into one appliance reading, simply add the L1 sub-meter and the L2 sub-meter readings together.

Each site can have one or more contiguous sampling blocks (\texttt{blk}). If there is a significant period of time where the capture of a house stops and then starts, we break that up into two blocks. This helps researchers and data scientists with algorithm testing where contiguous streams of time-series data are necessary. 
This data, along with other meta data (see Table~\ref{tbl:hmeta}), is stored in the ``\texttt{<type>?.txt}'' file. For~House 1, this file would be \texttt{house1.txt}.
Each block has the following files associated with it (\mbox{see Table~\ref{tbl:files}}). The \texttt{power} and \texttt{energy} files contain all real power measurements from mains and sub-meters (good for testing NILM). The \texttt{subs} files contain 11 electrical measurements for each sub-meter. When the HVAC system has electric heating and cooling, we include a \texttt{tstat} file that contains data from the house\rq{}s thermostat.

%%%%%%%%%%%%%%%%%%%%%%%%%%%%%%%%%%%%%%%%%%
% Main methods applied to collect and treat, as well as to use and reuse the data. Notes on validation and curation techniques applied. Notes on data quality, noise, etc.
\section{Methods}

When designing the data capture system for RAE, we prioritized the need for accuracy and reliability. Hence, we chose commercial-grade metering equipment. We chose to use the Rainforest Automation EMU2 in-home display\footnote{See \href{https://rainforestautomation.com/rfa-z105-2-emu-2/}{https://rainforestautomation.com/rfa-z105-2-emu-2/}.} to capture smart meter data.  See Table~\ref{tbl:power} (column name \texttt{ihd}) for the data we captured from the EMU2. The EMU2 reads data from a ZigBee-enabled smart meter at roughly 15 s intervals.

To capture sub-meter data, we chose a Class 1 branch circuit power meter from DENT, the~PowerScout 24\footnote{See \href{https://www.dentinstruments.com}{https://www.dentinstruments.com}.}. We had prior experience with using the DENT PowerScout 18 m. See Table~\ref{tbl:subs} for the data we captured from the PowerScout 24. The PowerScout 24 can monitor up to 24 circuits at a~rate of 1 Hz.

Thermostat data was collected from the EcoBee3 thermostat\footnote{See \href{https://www.ecobee.com}{https://www.ecobee.com}.} at 5 min intervals (a product limitation). Data includes set points, operation mode (heat/cool and stage), outdoor temperature and wind speed, and indoor humidity. Indoor temperature and motion is reported from the thermostat and three remote sensors (located in the living room, the basement rec room, and the master bedroom).

\begin{figure}[H]
\centering
\includegraphics[width=0.45\textwidth]{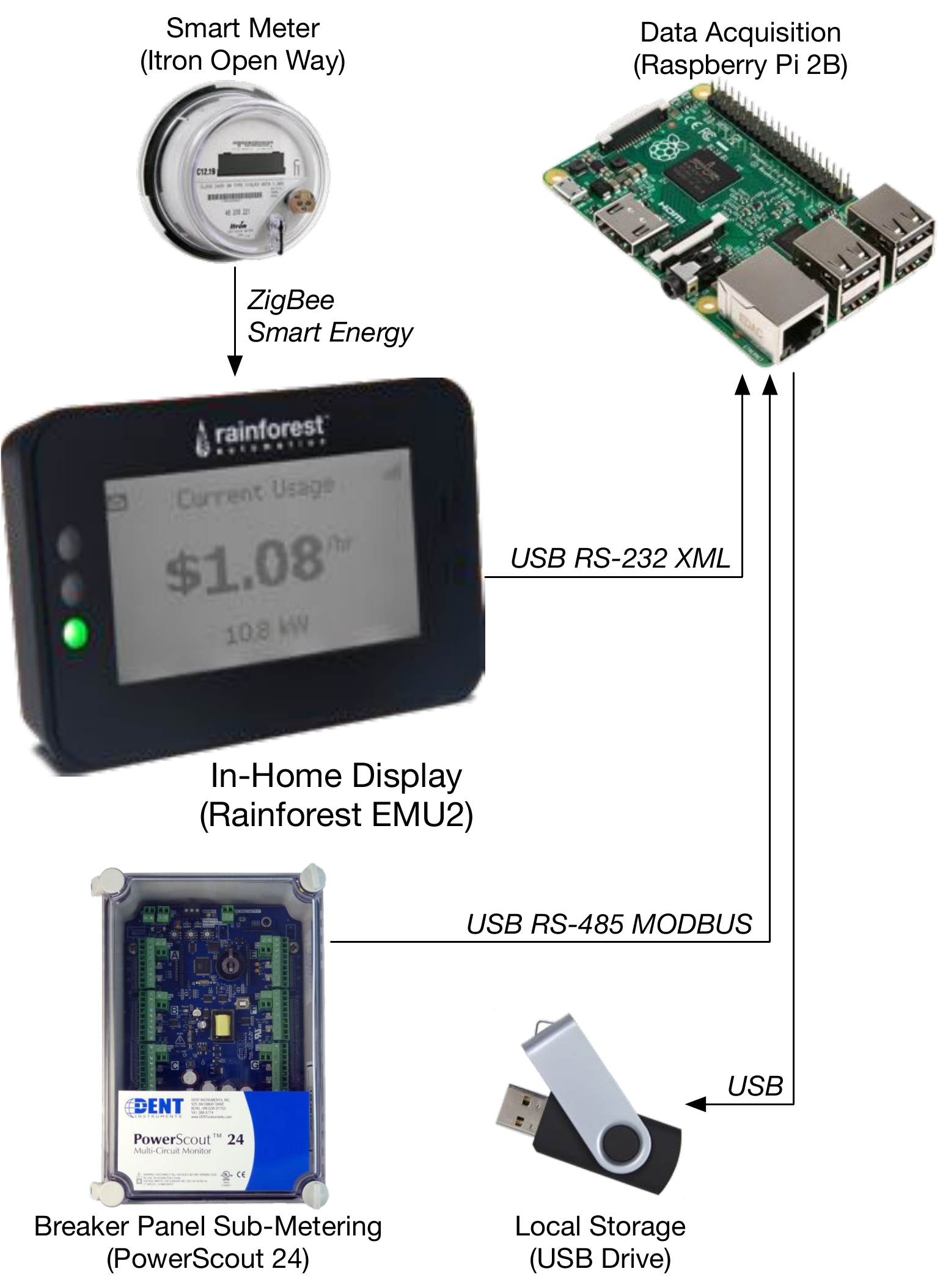}
\caption{Diagram of the data capturing hardware/setup.}
\label{fig:block}
\end{figure}

The hardware setup used to capture data for RAE is depicted in Figure~\ref{fig:block}, and we have released (as open source) the code\footnote{Code available on GitHub at \href{https://github.com/smakonin/RAEdataset}{https://github.com/smakonin/RAEdataset}.} used to capture, store, and convert the raw data. This setup is minimal and will allow us to easily install this equipment in a different house to capture data and add it to the RAE dataset.

Data that is missing will be represented by a timestamp and one or more null data-points. For \textit{comma-separated value} (CSV) files, this would mean no data between commas. For example, ``\texttt{1457282030,,,,4.582,38193.4}'' would mean that three readings are missing.

%%%%%%%%%%%%%%%%%%%%%%%%%%%%%%%%%%%%%%%%%%
% Further notes on the usage of the data set that will help other researchers to access and further understand practical aspects of working with the data. If there are ethical or compelling commercial reasons that the data cannot be made available, either in part or in full, these should be described in as much detail as possible. You should make clear how the data can be accessed and if there are circumstances in which access would be denied (e.g. if complying with the request would compromise anonymity of human participants or if an embargo applies); we recommend full access wherever possible.
\section{Usage Notes}

\subsection{House 1 Energy Consumption Analysis}

\begin{figure}[H]
\centering
\includegraphics[width=0.65\textwidth]{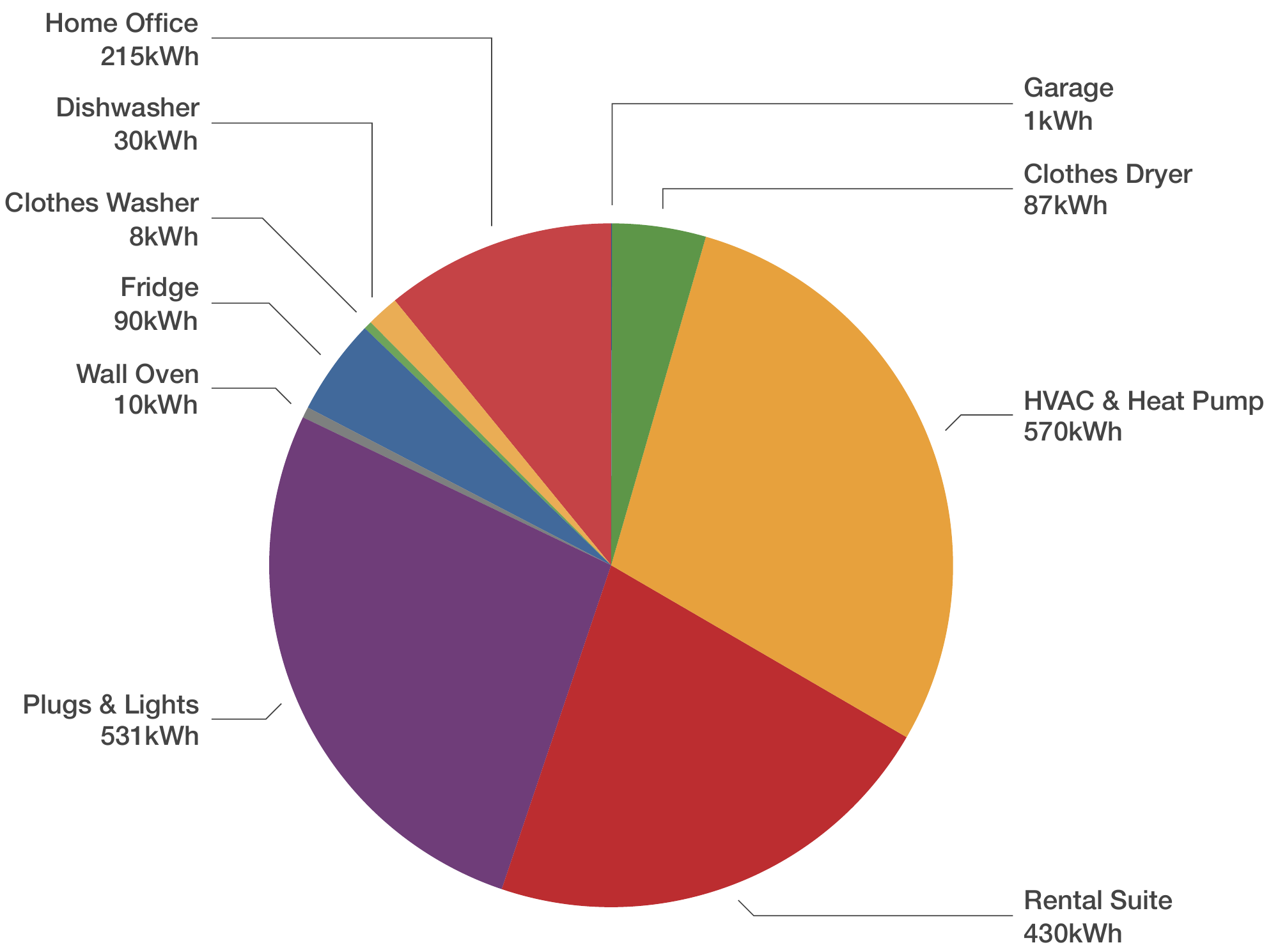}
\caption{Percentages of energy consumed (in kWh) over the 72-day period for a total of 1971 kWh.}
\label{fig:pieh1}
\end{figure}

The three highest consumers of energy in House 1 were the \textit{HVAC \& Heat Pump} (570 kWh), \textit{Plugs~\&~Lights} (531 kWh), and \textit{Rental Suite} (430 kWh), as shown in Figure~\ref{fig:pieh1}. Over the 72-day capture period, the smart meter reported a total energy consumption of 1982 kWh. A total of 1971 kWh was found when each of the 24 sub-meters real energy accumulator is summed up. There is an 11 kWh discrepancy due to the rounding errors in each sub-meter accumulator as each sub-meter reports only whole-Watt measurements. Additionally, the smart meter from the utility is a Class 1 m, whereas the sub-meters are Class 0.5. This means there is a higher measurement error in the readings from the smart meter.

\subsection{House 2 Energy Consumption Analysis}

\begin{figure}[H]
\centering
\includegraphics[width=0.65\textwidth]{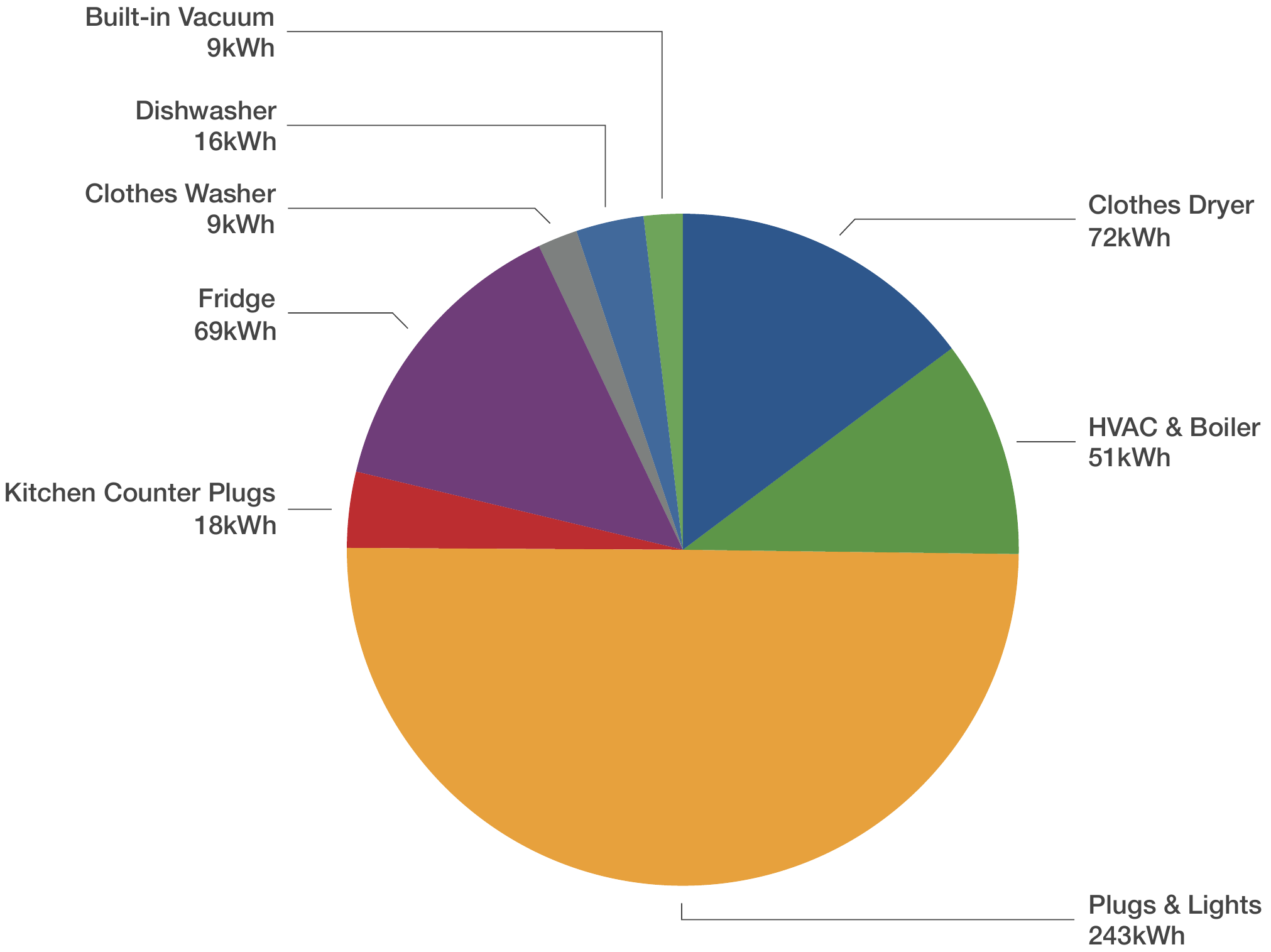}
\caption{Percentages of energy consumed (in kWh) over the 59-day period for a total of 478 kWh.}
\label{fig:pieh2}
\end{figure}

House 2 is a smaller (26.1 m$^2$ less space) and more energy-efficient house than House 1. \textit{Plugs \& Lights} (242.5 kWh) were the highest consumers of energy, as shown in Figure~\ref{fig:pieh2}. Over the 59-day capture period, the smart meter reported a total energy consumption of 478 kWh. A total of 497 kWh is found when each of the 21 sub-meters real energy accumulator is summed up. There is a 19 kWh discrepancy which is due to the same issues mentioned in the previous sub-section.

\vspace{-12pt}

\subsection{NILM Example}

We wanted to use the RAE dataset to test the accuracy of the NILM algorithm. 
For this, we used the SparseNILM algorithm~\cite{makonin2016tsg}. 
SparseNILM uses a variant of the Viterbi algorithm to find the most likely set of appliances that are ON in each time period (as well as their power level) and a rate matching the dataset used --- in this case, 1 Hz.
We ran our test on a MacBook Pro (13-inch, Late 2016) having a 3.3 GHz Intel Core i7 processor with a 16 GB memory.

First, we removed the rental suite sub-panel power data so that we could test for a single occupancy home.
Second, we picked six high-consuming loads (clothes dryer, furnace, heat pump, oven, fridge, and dishwasher) to disaggregate. 
Third, we trained the algorithm using data from the first block file (nine days). 
This resulted in the creation of a 2000-state hidden Markov model (HMM) that modeled all six loads. 
The training phase (consisting of one iteration) took 58 s to complete.

\begin{figure}[H]
\centering
\includegraphics[width=0.6\textwidth]{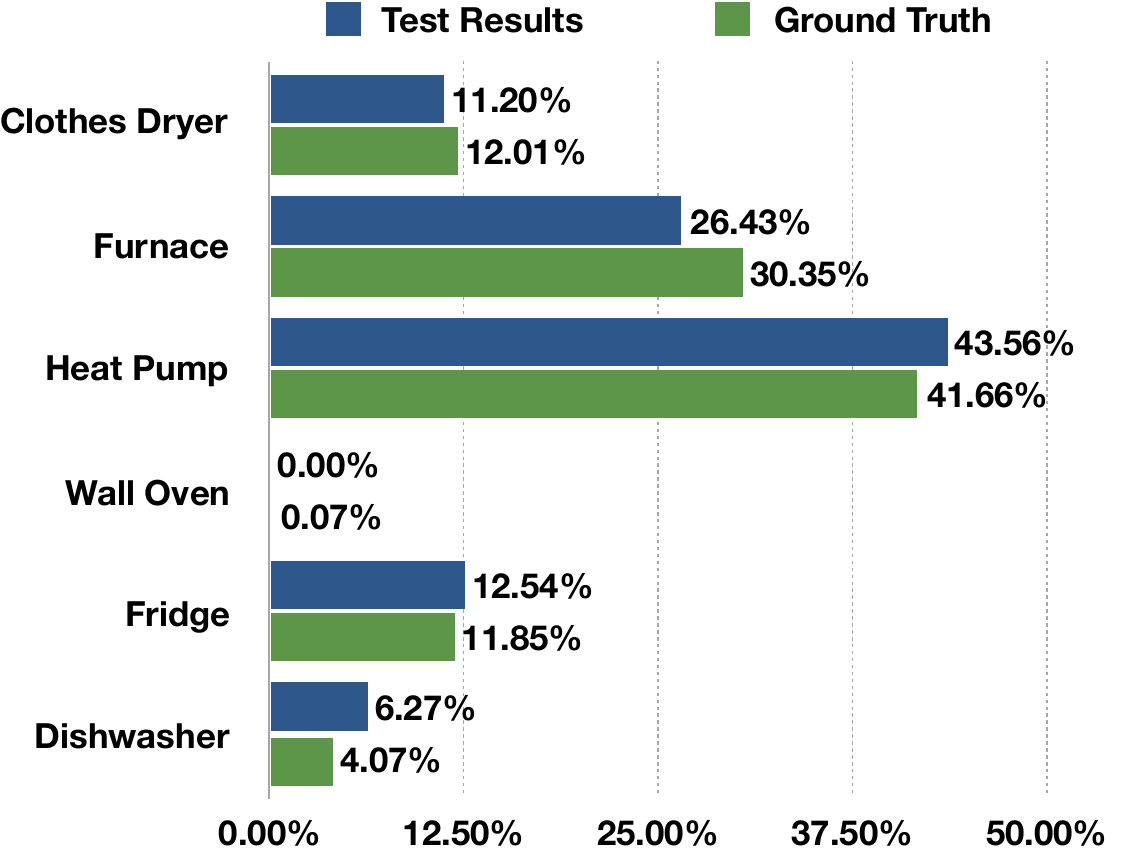}
\caption{Appliance/load-specific accuracy results (in percentages of total desegregated, not of the total house).}
\label{fig:bar}
\end{figure}

Next, we tested the accuracy of our HMM by having it disaggregate the data from the second block file (63 days). 
Testing took 46 min to complete, disaggregating 5.4 million samples with an average disaggregation time of 330 $\upmu$s per sample,. 
We report overall accuracy  results in Table~\ref{tbl:results}. 
Figure~\ref{fig:bar} shows the accuracy results of each appliance/load that was disaggregated.
Our experiment yielded an accuracy of over 80\% and very low error results.

\vspace{-12pt}

\begin{table}[H]
  \caption{Overall accuracy results of our NILM test.}
  \centering
  \begin{tabular}{cc}
  \toprule
	\textbf{Accuracy Metric} 												& \textbf{Score}    \\
  \midrule
	Precision 														&  87.86\% \\
	Recall 															&  85.01\% \\
	F-score 														&  86.41\% \\
	Finite-State F-score (FS-fscore)~\cite{makonin2014eval} 			&  80.47\% \\
	Normalized Disaggregation Error (NDE)~\cite{parson2012non}     	&   0.71\% \\
	Root-Mean-Square Error (RMSE)									&  62.14   \\
  \bottomrule
  \end{tabular}
  \label{tbl:results}
\end{table}

%%%%%%%%%%%%%%%%%%%%%%%%%%%%%%%%%%%%%%%%%%
\vspace{6pt}

%%%%%%%%%%%%%%%%%%%%%%%%%%%%%%%%%%%%%%%%%%
%% optional
%\supplementary{The following are available online at www.mdpi.com/link, Figure S1: title, Table S1: title, Video S1: title.}

%%%%%%%%%%%%%%%%%%%%%%%%%%%%%%%%%%%%%%%%%%
\acknowledgments{This work was funded in part by an NSERC Engage Grant EGP-501582-16.}
%All sources of funding of the study should be disclosed. Please clearly indicate grants that you have received in support of your research work. Clearly state if you received funds for covering the costs to publish in open access.}

%%%%%%%%%%%%%%%%%%%%%%%%%%%%%%%%%%%%%%%%%%
\authorcontributions{S.M. conceived and designed the data capturing systems and is the main author. Z.J.W.~provided supervision as well as manuscript feedback and editing. C.T. provided support for the Embedded Automation hardware, guidance, manuscript feedback, and editing.}
%For research articles with several authors, a short paragraph specifying their individual contributions must be provided. The following statements should be used ``X.X. and Y.Y. conceived and designed the experiments; X.X. performed the experiments; X.X. and Y.Y. analyzed the data; W.W. contributed reagents/materials/analysis tools; Y.Y. wrote the paper.'' Authorship must be limited to those who have contributed substantially to the work reported.}

%%%%%%%%%%%%%%%%%%%%%%%%%%%%%%%%%%%%%%%%%%
\conflictsofinterest{The authors declare no conflict of interest. The founding sponsors had no role in the design of the study; in the collection, analyses, or interpretation of data; in the writing of the manuscript; or in the decision to publish the results.}
\reftitle{References}

% The following MDPI journals use author-date citation: Arts, Econometrics, Economies, Genealogy, Humanities, IJFS, JRFM, Laws, Religions, Risks, Social Sciences. For those journals, please follow the formatting guidelines on http://www.mdpi.com/authors/references
% To cite two works by the same author: \citeauthor{ref-journal-1a} (\citeyear{ref-journal-1a}, \citeyear{ref-journal-1b}). This produces: Whittaker (1967, 1975)
% To cite two works by the same author with specific pages: \citeauthor{ref-journal-3a} (\citeyear{ref-journal-3a}, p. 328; \citeyear{ref-journal-3b}, p.475). This produces: Wong (1999, p. 328; 2000, p. 475)

%=====================================
% References, variant B: external bibliography
%=====================================
%\externalbibliography{yes}
%\bibliography{z-refs}

%%%%%%%%%%%%%%%%%%%%%%%%%%%%%%%%%%%%%%%%%%
%% optional
%\sampleavailability{Samples of the compounds ...... are available from the authors.}

%%%%%%%%%%%%%%%%%%%%%%%%%%%%%%%%%%%%%%%%%%
\end{document}